\documentclass[a4paper,12pt]{iopart}

\expandafter\let\csname equation*\endcsname\relax
\expandafter\let\csname endequation*\endcsname\relax

\usepackage{amsmath}
\usepackage{hyperref}
\usepackage{cite}
\usepackage{graphicx}
\usepackage{amsfonts}
\usepackage{amsthm}
\usepackage{cases}
\usepackage{bm}
\usepackage{subcaption}

\usepackage{fixmath}

\usepackage{enumitem} 
\usepackage[normalem]{ulem}

\usepackage{color}
\definecolor{Blue}{rgb}{0.0, 0.0, 0.8}
\definecolor{Red}{rgb}{0.80, 0.00, 0.00}
\definecolor{Green}{rgb}{0.00, 0.50, 0.00}
\definecolor{Black}{rgb}{0.0, 0.0, 0.0}

\newcommand{\blue}{\color{Black}}

\hypersetup{
    colorlinks=true,       
    linkcolor=red,          
    citecolor=blue,        
    filecolor=magenta,      
    urlcolor=cyan           
}

\newcommand{\be}{\begin{equation}}
\newcommand{\ee}{\end{equation}}
\newcommand{\bea}{\begin{eqnarray}}
\newcommand{\eea}{\end{eqnarray}}

\newcommand{\beq}{\begin{equation}}
\newcommand{\eeq}{\end{equation}}
\newcommand{\beqn}{\begin{eqnarray}}
\newcommand{\eeqn}{\end{eqnarray}}

\newcommand{\abs}[1]{\ensuremath{\left| #1 \right|}}

\begin{document}

\title{The cost of resetting discrete-time random walks}

\author{John C. Sunil$^1$\footnote{Corresponding Author}, Richard A. Blythe$^1$, Martin R. Evans$^1$ and Satya N. Majumdar$^2$}
\address{$^1$ SUPA, School of Physics and Astronomy, University of Edinburgh, Peter Guthrie Tait Road, Edinburgh EH9 3FD, UK}
\address{$^2$ Universit{\'e} Paris-Saclay, CNRS, LPTMS, 91405, Orsay, France}
\eads{\mailto{j.chakanal-sunil@sms.ed.ac.uk}, \mailto{r.a.blythe@ed.ac.uk}, \mailto{m.evans@ed.ac.uk}, \mailto{satya.majumdar@universite-paris-saclay.fr}}

\begin{abstract}
We consider  a discrete-time continuous-space random walk, with a symmetric jump distribution, under stochastic resetting.  {\blue Our aim is to understand how costs, associated with the random walk jumps  and the resetting, contribute to the total cost of the motion. We calculate the distribution of the total cost up to first passage of the random walker to a target.} By using the backward master equation approach we demonstrate that the distribution of the total cost  can be reduced to a Wiener-Hopf integral equation. The resulting integral equation can be exactly solved (in Laplace space) for arbitrary cost functions for the jump and selected functions for the reset cost. We show that the large cost behaviour is dominated by resetting or the jump distribution according to the choice of the jump distribution. In the important case of a Laplace jump distribution, which corresponds to run-and-tumble particle dynamics, and linear costs for jumps and resetting,
the Wiener-Hopf equation simplifies to a differential equation
which can easily be solved.

\end{abstract}

\today

\submitto{\JSTAT}

\maketitle

\section{Introduction}

Over time, a physical entity may undergo motions of several different types. For example, a foraging animal \cite{Stanley,Bell} might spend time searching a local area for food, and occasionally traverse greater distances more rapidly to locate a more fertile patch. Different costs may be associated with these different motions, for example, an energy cost with the local moves and the risk of moving to a unknown area with the longer-range moves. The effect of combining a `normal' motion (such as diffusion  or run and tumble dynamics) with stochastic \emph{resetting} to an initial or other fixed position has attracted intense scrutiny \cite{EM11a,EM11b,KMSS14,EM18,EMS20}, wherein the reset has a cost that may be a time penalty \cite{RUK14,EM19,BS20a,BS20b,PKR20,SBEM23,SBEM24,ES24}, energetic cost \cite{TPSRR20,BBPMC20,FBPCM21,BP21,GLPHMG23b,OGMK24,TKRR24,Singh2024} or entropy production cost \cite{FGS16,MOK23}.  Here, our aim is to understand how costs associated with the normal motion and the resetting combine, with a particular focus on properties up to the first passage to a pre-determined position \cite{Redner,Majumdar07,SM14,MM20,SP22} and identifying regimes where one cost dominates. These types of first passage problems model scenarios such as a foraging animal successfully finding food, or the decision to buy or sell in financial markets.

We pursue our aim in the context of a random walker in discrete time and continuous space, where the normal motion comprises jumps drawn from a general distribution, and a reset to the initial position occurs with a constant probability at each time step. More precisely, we denote the position of the random walker at the $k^{\text{th}}$ time step as $x_k$. The random walker jumps to the new position $x_k + \eta_k$ with a probability $(1-r)$ with $\eta$ drawn from a jump distribution $f(\eta)$.  With complementary probability $r$ the random walker is reset to position $x_r$. Resetting in discrete-time jump processes, without costs,  was  studied in \cite{KMSS14}. Here we introduce  two separate cost functions for the jumps and the resets, both of which  depend on the distance travelled by the particle: for jumps the cost is $h(\eta_k)$ and for resets the cost is  $g(x_r-x_{k-1})$.  We note that the jump dynamics is very general, encompassing, for example, standard diffusion and L\'evy flights as special cases. As such, we extend separate studies of the cost of diffusion with resetting \cite{SBEM23,SBEM24,PKR20,BS20a,MOK23,GC25} and the cost of random walk dynamics without resetting \cite{BCP22,MMV23a,MMV23b,MMV24}. In particular, \cite{SBEM23} considered diffusion for a general class of resetting costs, while \cite{MMV24} found a general formula for the cost up to first passage to the target for a general jump process with a general cost function but no resetting. 
The model considered here thus combines the disparate features of \cite{SBEM23} and \cite{MMV24} to obtain a model of a discrete-time jump process with resetting.  

The discrete-time jump process may also be mapped onto the motion of a run and tumble particle in continuous time, as noted in \cite{MMV24}. A run and tumble particle models an active entity such as  a bacterium that consumes energy from the environment and moves via self-propulsion \cite{TC08}. The particle performs a run, in its direction of motion, whose length  is a random variable before tumbling and choosing anew its direction of motion. In the mapping onto a discrete-time jump process the run length becomes the jump length of the discrete-time process and the discrete times  become the tumbling times. In one dimension a symmetric jump length distribution corresponds to the run and tumble particle choosing the forward or backward directions with equal probability. The standard run and tumble particle, also known as a persistent random walker, moves with constant speed $v_0$ and the tumbles occur at constant rate so that the time between tumbles is exponentially distributed. This amounts to a Laplace distribution for the jumps $f(\eta) = (\lambda /2) \exp(-\lambda |\eta|)$ where $\lambda = \gamma/v_0$  and $\gamma$ is the tumbling rate. Resetting  of run and tumble dynamics has been of particular interest \cite{EM18,masoliver19,SBS20,Bressloff20,PPPPL24} and  it is  natural to consider a resetting cost for this process.

We study the  general discrete-time jump process  described in the preceding paragraphs for two different choices of reset cost functions: constant cost and linear cost, with arbitrary jump distributions and jump costs. The constant reset cost plays the  role of a refractory period after each reset\cite{RUK14,EM19}, whereas a linear cost plays the role of the time required to reset the particle back to the starting position with a constant velocity\cite{BS20a,PKR20,SBEM23,SBEM24}. Our main findings take the form of exact closed form expressions in Laplace space for the total cost up to first passage to the target for each of the two reset cost functions. Further we also compute  how the cost behaves in the asymptotic regime.  
In the case of constant resetting cost,  our results reveal that 
for  jump cost  distributions with finite mean cost per jump, resetting converts the power-law distribution of large costs observed in \cite{MMV24} to an exponentially distributed tail. On the other hand, when the mean cost per jump is not finite, the jumps dominate and the power-law tail is retained. 

The paper is organised as follows, in Section~\ref{sec:mod_def}, we define the model of a discrete time random walker under stochastic resetting, using which we derive the backward master equation (BME) for the total cost up to first passage to the target. We then present the important exactly solvable case of a Laplace jump distribution (negative exponential of the magnitude of the jump) 
and costs that are linear in the distance travelled
in Section~\ref{sec:exact_sol}.  As outlined  above, this case is of particular importance as it describes, for example, run and tumble particles.
 In this case the  backward equation may be reduced to a differential equation that can be solved by elementary methods
and a closed form for the Laplace transform of the cost distribution is obtained.
In Section~\ref{sec:gensol} we review how in the case of a general jump distribution the backward equation
reduces to a Wiener-Hopf integral equation with inhomogeneous terms.
The solution of the Wiener-Hopf integral equation is in general a difficult task, but we show that, in our case of constant or linear resetting costs,  the solution may be constructed
from known solvable cases of the Wiener-Hopf equation.
We  present the closed form results for a few simplifying
choices of the jump distribution and jump cost distribution in Sections~\ref{sec:const_cost} and ~\ref{sec:lin_cost} and discuss the asymptotic behaviours in Section~\ref{sec:asymptotics}. 
Finally, we conclude and summarise in Section~\ref{sec:conclusion}. 

\section{Set-up of the Problem} \label{sec:mod_def}
\begin{figure}
    \centering
    \includegraphics[width = 0.6\textwidth]{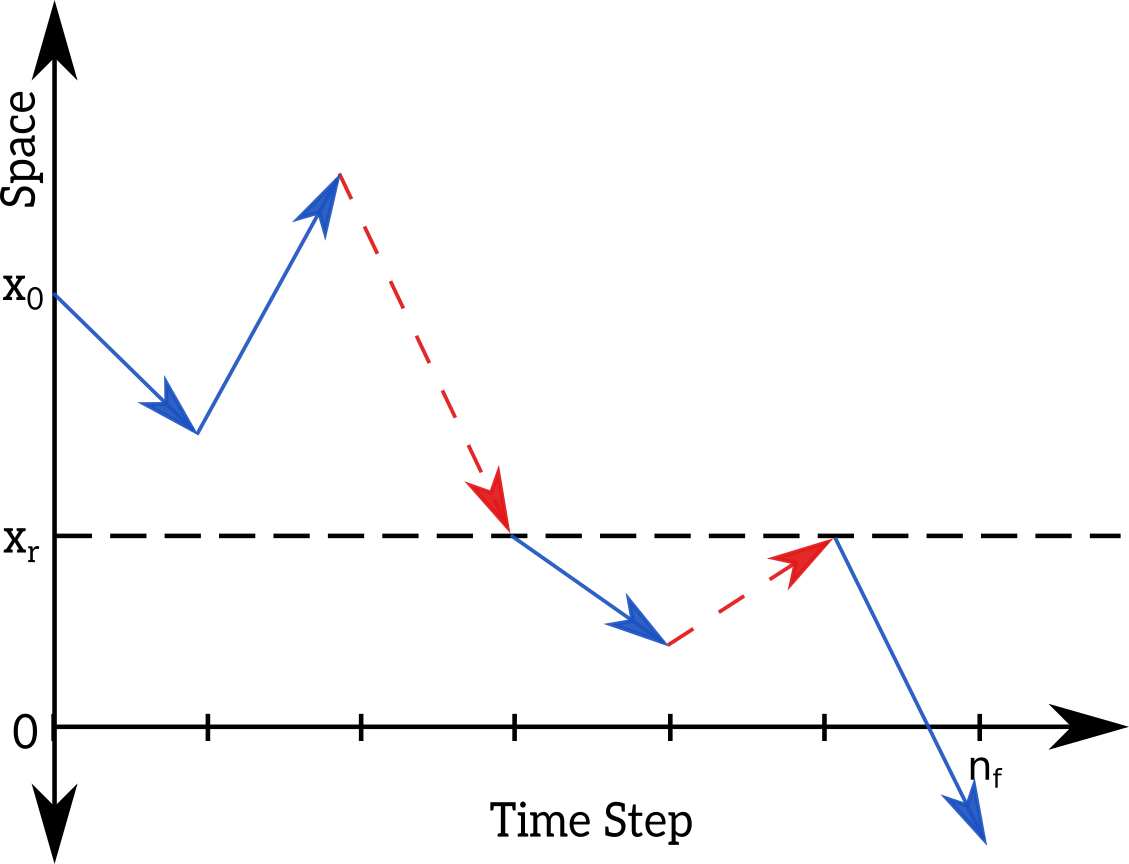}
    \caption{Schematic of discrete time resetting jump process that is terminated upon first passage to the origin. The red dashed lines represent the resets that occur for the given trajectory.}
    \label{fig:schem}
\end{figure}

We consider a discrete time resetting process \cite{EMS20,KMSS14} where the particle starts at $x_0$ and resets to $x_r$ with probability $r$. If the particle does not reset in a given time step, it jumps a distance drawn from the symmetric, continuous and normalized jump distribution $f(\eta)$. Hence, the evolution of the position of the particle, $x_k$, at time step $k$ is given by
\begin{align}
   x_k = 
   \begin{cases} 
      x_{k-1} + \eta_k & \text{with probability $1-r$}\;, \\
      x_r & \text{with probability $r$} \;,
   \end{cases}
\end{align}
with initial position $x_0$.
We focus on the first passage properties of this system. The process is terminated when the random walker crosses the origin at time step $n_f$. A schematic of a sample trajectory is given in Fig.~\ref{fig:schem}.
To each jump we associate a cost $h(\eta)$ and with each reset we associate a cost $g(\eta)$ where $\eta$ is the distance traversed in the jump or reset. For the moment, we only impose that the cost function is positive ($h(\eta) \geq 0, g(\eta) \geq 0$) and symmetric ($h(\eta) = h(-\eta), g(\eta) = g(-\eta)$). Thus the total cost for the process evolves as,
\begin{align}
   C_k = 
   \begin{cases} 
      C_{k-1} + h(\eta_k) & \text{with probability $1-r$}\;, \\
      C_{k-1} + g(x_r-x_{k-1}) & \text{with probability $r$} \;.
   \end{cases}
\end{align}
We are interested in the cost that the process incurs up to the first passage to the origin, which occurs when $x_k<0$ for the first time. The cost of similar discrete-time processes without resetting has been studied in \cite{MMV23a,MMV23b,BCP22}. {\blue
The costs, $g(\eta)$ and $h(\eta)$ could be interpreted as the time required to implement the respective motions, the monetary costs associated or the energy consumed in undergoing the motions of jumps and resets. Depending on the physical context $g(\eta)$ and $h(\eta)$ will have various particular forms.
To begin with,  we  keep $g(\eta)$ and $h(\eta)$ to be general symmetric functions and specialise to the cases of constant or linear $g(\eta)$ in Sections \ref{sec:const_cost} and \ref{sec:lin_cost}. 
}

{\blue First passage problems for discrete-time random walkers can be extended to asymmetric jump distributions, which imply a drift or bias in the jump motion. The first passage properties of such asymmetric jump distributions have been studied in \cite{Majumdar10}.  Furthermore, dropping the requirement that $g(\eta)$ and $h(\eta)$ are even leads us to anisotropic costs where jumping or resetting in one particular direction has a different cost. In what follows, we restrict the problem to symmetric jumps and costs for simplicity. }

\subsection{Backward Master Equation}
The quantity of interest is  $Q_r(x_0,C)$, which is  the distribution of the total cost for a random walker up to the first passage to the origin, given that it starts at $x_0$ and resets to $x_r$ with probability $r$. Note that $Q_r(x_0,C)$ implicitly depends on the resetting position $x_r$, but we do not display the dependence here for  brevity of notation. In order to calculate such distributions, it is natural to use a backward master equation approach. Using the Markov property for the process, we can write down the  backward master equation as an integral equation
\begin{align}
    Q_r(x_0,C) = &(1-r)\Bigg[ \int_0^\infty dC^\prime \int_0^\infty dx_1 Q_r(x_1,C^\prime)f(x_1-x_0)\delta(C^\prime+h(x_1-x_0)-C) \nonumber\\ &\qquad \qquad+ \int_{-\infty}^{0^{-}} dx_1 \delta(C-h(x_1-x_0))f(x_1-x_0)\Bigg] \nonumber \\& + r \Bigg[ \int_0^\infty dC^\prime Q_r(x_r,C^\prime)  \delta(C^\prime + g(x_r-x_0)-C) \Bigg]\;. \label{int_FPE}
\end{align}
The first term in \eqref{int_FPE} corresponds to the event when there is no resetting in the first time step, which occurs with probability $(1-r)$, while the second term corresponds to the case when the particle undergoes a reset at the very first step, which happens with probability $r$. The first part of the first term in \eqref{int_FPE} corresponds to an initial jump where the particle ends up at a position $x_1>0$, after which the process repeats again, starting from $x_1$. The second part accounts for an initial jump which takes the particle to a position $x_1<0$. There is a cost associated with this jump, but the process is then terminated. {\blue The upper limit of this integral is $0^-$ as the process does not terminate if the random walker is exactly at $0$. However, this does not affect the calculations as we later assume $f(x_1-x_0)$ to be a continuous function.} Note that the process is not immediately terminated on the first passage to the origin, but the random walker is allowed to finish the complete step and the cost is accounted for the complete step, as shown in Fig.~\ref{fig:schem}. {\blue This choice of including the cost associated with the leapover length is for analytical simplicity. A detailed analysis of the distribution of the leapover length for various kinds of L\'evy flights is presented in \cite{KLCKM07}.} The final possibility at the initial step is that the particle resets to the resetting position, $x_r$, and this is accounted  for by the last term in \eqref{int_FPE}.

We define the Laplace transform of \eqref{int_FPE} with respect to the total cost as
\begin{align}
    \widetilde{Q}_r(x_0,p) = \int_0^{\infty}dC\; e^{-pC}Q_r(x_0,C)\;.
\end{align}
Taking the Laplace transform of \eqref{int_FPE} we obtain the  following integral equation,
\begin{align}
    \widetilde{Q}_r(x_0,p) = &(1-r)\Bigg[ \int_0^\infty dx_1 \widetilde{Q}_r(x_1,p)e^{-p h(x_1-x_0)}f(x_1-x_0)\nonumber + \\ &\int_{-\infty}^{0^{-}} dx_1 e^{-p h(x_1-x_0)}  f(x_1-x_0) \Bigg] + r  e^{-p g(x_r - x_0)} \widetilde{Q}_r(x_r,p) \;. \label{LT_FPE}
\end{align}
The problem now is to solve equation \eqref{LT_FPE} to obtain the Laplace transform of the total cost distribution up to first passage to the origin.

\section{Exactly Solvable Case: Laplace Jump Distribution with Linear Costs} \label{section_exact_solution} \label{sec:exact_sol}

The integral equation \eqref{LT_FPE} is quite difficult to solve in general, but for a particular choice of jump length distribution and cost functions, it can be reduced to a differential equation, which can then be easily solved. 
In this way the Laplace jump distribution with linear costs provides a case which may be solved exactly in a self-contained way.

We consider a Laplace distribution for the jump lengths $f(\eta) = \frac{1}{2}e^{-\abs{\eta}}$, and linear costs for jumps and resets, $ h(\eta) = c_j\abs{\eta}$ and $g(\eta) = c_r\abs{\eta}$. As explained in the introduction, the Laplace jump distribution
furnishes a description of a `run and tumble particle' which is a persistent random walker that continues moving in the same direction for exponentially distributed time intervals before `tumbling' i.e. switching to another direction \cite{ACEDK14,pozzoli23,MMV24}.
The linear cost  models the consumption of some resource in proportion to the distance travelled between tumbles.

With these choices, we obtain from \eqref{LT_FPE}
\begin{align}
    \widetilde{Q}_r(x_0,p) = &(1-r)\left[\int_0^\infty dx_1 \widetilde{Q}_r(x_1,p)\frac{1}{2}e^{-(p c_j+1) \abs{x_1-x_0}} + \int_{-\infty}^{0^-} dx_1 \frac{1}{2}e^{-(p c_j+1) \abs{x_1-x_0}} \right] \nonumber \\ &+ r  e^{-p c_r \abs{x_r - x_0}} \widetilde{Q}_r(x_r,p)\;. \label{Laplace_FPE}
\end{align}
The integral equation in \eqref{Laplace_FPE} can be converted to a differential equation by making use of the following identity,
\begin{align}
    \frac{d^2}{dx^2} e^{-a\abs{x-b}} = -2a\delta(x-b) + a^2 e^{-a\abs{x-b}}\;.
\end{align}
Then, taking the second derivative with respect to $x_0$ and integrating over the delta functions, we obtain from \eqref{Laplace_FPE} the following differential equation,
\begin{align}
    \frac{d^2}{dx_0^2} \widetilde{Q}_r(x_0,p) = &(p c_j +1) (p c_j  + r) \widetilde{Q}_r(x_0,p) + r\Big[ (p c_r)^2 e^{-p c_r \abs{x_r - x_0}} \nonumber \\ &-(p c_j + 1)^2e^{-p c_r \abs{x_r-x_0}} - 2p c_r \delta(x_r - x_0) \Big]\widetilde{Q}_r(x_r,p)\;. \label{Laplace_DE}
\end{align}
In \eqref{Laplace_DE}, we have an inhomogeneous term, $\widetilde{Q}_r(x_r,p)$, that itself depends on the function to be solved,  so the equation has to be solved self-consistently.
This method of taking the second derivative to convert an integral equation to a differential equation has been used previously in \cite{CM05,MS24}.

The general  solution to \eqref{Laplace_DE}  for all $x_0$ and $x_r$ is presented in \ref{appendix_sol_DE}. As the general solution is cumbersome, here we look at the special case of $x_0 = x_r = 0$, which is the case where the particle both starts and resets to the origin. We then obtain the final result,
\begin{multline}
    \widetilde{Q}_r(0,p) =  \\
    \frac{p \left(p c_j^2+r c_j+c_j-p c_r^2\right)+r}{p \left(p^2 c_j^3+p c_j^2 (\mu +r+2)+c_j \left(p c_r \left(r-p c_r\right)+\mu +2 r+1\right)+c_r \left(r-(\mu +1) p
   c_r\right)\right)+r}\;, \label{Qr_gen}
\end{multline}
where $\mu = \sqrt{(1+pc_j)(r+pc_j)}$.
Expression \eqref{Qr_gen} is the main result of this section.
It simplifies considerably for certain choices of $c_j$, $c_r$, which we now discuss.

\subsection{Reset cost only}
If only resets have cost, $c_j = 0, c_r = 1$, we have
\begin{align}
     \widetilde{Q}_r(0,p) = \frac{p+\sqrt{r}}{p \sqrt{r}+p+\sqrt{r}}\;,
     \label{Laplace_cost_dist01}
\end{align}
This expression for the Laplace transform  may be inverted exactly to obtain
\begin{equation}
    Q_r(0,C) = \frac{1}{1+r^{1/2}} \delta(C) +  \frac{r}{(1+r^{1/2})^2}e^{-Cr^{1/2}/(1+r^{1/2})}\;.
\end{equation}
We see there is a probability $1/(1+r^{1/2})$ of zero cost which corresponds to first passage to the origin without any resets. The other piece of the distribution  is an exponential distribution with
characteristic cost $1+r^{-1/2}$, which diverges as $r\to 0$.

\subsection{Jump cost only}
If only jumps have cost,  $c_j = 1, c_r = 0$  we have
\begin{align}
     \widetilde{Q}_r(0,p) = \frac{(p+1)(p+r)-p\sqrt{(p+1)(p+r)}}{(p+1)(p+r+pr)}\;. \label{Laplace_cost_dist10}
\end{align}
This Laplace transform is difficult to invert exactly, but we can obtain the asymptotic large $C$ behaviour from the pole at $p=-r/(1+r)$:
\begin{equation}
    Q_r(0,C) \sim  \frac{2r^2}{(1+r)^2}e^{-Cr/(1+r)}\;.
\end{equation}
Thus the large cost distribution is exponential with
characteristic cost $1+r^{-1}$, which has a stronger divergence as $r\to 0$ than the reset-only distribution.

Finally we note  that plugging in $r = 0$ in \eqref{Laplace_cost_dist10} recovers the result for the jump cost distribution without resetting as obtained in \cite{MMV24}
\begin{align}
     \widetilde{Q}_0(0,p) = 1-\sqrt{\frac{p}{1+p}}\;.
\end{align}
An exact inversion of this  Laplace transform was found in \cite{MMV24}, and the asymptotic behaviour  determined to be a power law 
\begin{equation}
    Q_0(0,C) \sim  \frac{1}{\sqrt{4\pi}}\frac{1}{C^{3/2}}\;.\label{pl}
\end{equation}

\subsection{Jump and reset cost}
\begin{figure}
    \centering
    \includegraphics[width = 0.6\textwidth]{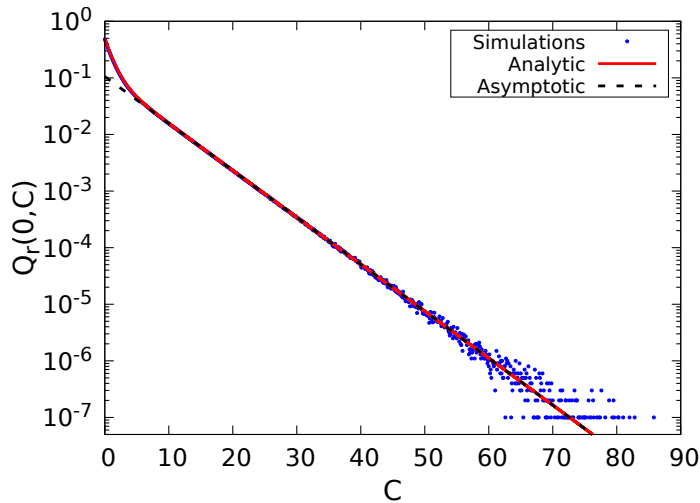}
    \caption{Distribution of cost obtained for Laplace jump distribution with linear jump and resetting costs for $r = 0.5$. Note the logarithmic y-axis, indicating an exponential asymptotic behaviour. }
    \label{fig:Laplace_all_costs}
\end{figure}

If the jump and reset costs are weighted equally, $c_j = c_r = 1$, we obtain from \eqref{Qr_gen},

\begin{align}
    \widetilde{Q}_r(0,p)= \frac{\sqrt{1+p}(r+ (2r+1)p)-p \sqrt{p+r}}{\sqrt{1+p} (4rp^2+(4r+1)p+r)}\;. \label{Laplace_cost_dist11}
\end{align}
Inverting \eqref{Laplace_cost_dist11} numerically \cite{AW06}, we obtain the distribution of the total cost at first passage to the origin with $x_0=x_r=0$,  plotted in Fig.~\ref{fig:Laplace_all_costs}. The asymptotic behaviour for large costs can again be obtained by finding the largest singularity of \eqref{Laplace_cost_dist11}. There are four singularities: two branch points at $p = -1,-r$ and two simple poles at $p_{\pm}$ given by 
\begin{align}
    p_{\pm} = \frac{-1-4r\pm\sqrt{1+8r}}{8r}\;.
\end{align}
 The largest singularity of \eqref{Laplace_cost_dist11} is then found to be the pole at $p_+$ and from this pole we obtain the leading large $C$  behaviour as an exponential decay
 \begin{align}
     Q_r(0,C) \sim \exp\left[-\left(\frac{{(4r+1)-\sqrt{1+8r}}}{8r}\right) C\right]\;. 
 \end{align}
 For $r=0.5$, we obtain {\blue from this expression (including the prefactor)} $ Q_r(0,C) \sim 0.1056 e^{-0.191C}$, which is plotted in Fig.~\ref{fig:Laplace_all_costs}. We observe again that 
the large cost behaviour is an exponential decay.

Thus  for a Laplace jump distribution with linear costs, in all three cases (reset cost, jump cost, jump and reset cost) we have seen that
the presence of resetting changes the large tail behaviour 
of the cost distribution  from a power-law decay \eqref{pl} in the absence of resetting, to an exponential decay.
\section{General Jump Distribution} \label{sec:gensol}
 In the case of general jump distribution $f(\eta)$ we must
 return to the backward equation \eqref{LT_FPE}.
To convert \eqref{LT_FPE} to a known integral equation called the Wiener-Hopf integral equation \cite{Ivanov94,Majumdar10,MMV24}, we normalise the kernel of the integrals, $e^{-p(\eta)}f(\eta)$,  by defining,
\begin{align}
    \chi_p(\eta) = \frac{e^{-ph(\eta)} f(\eta)}{2 A(p)}
\end{align}
where
\begin{align}
     \quad A(p) = \int_0^{\infty}d\eta\;e^{-ph(\eta)}f(\eta)\;, \label{A_def}
 \end{align}
using which we can re-write \eqref{LT_FPE} as
\begin{multline}
     \widetilde{Q}_r(x_0,p) -2(1-r)A(p)\int_0^\infty dx_1\; \widetilde{Q}_r(x_1,p)\chi_p(x_1-x_0)\\ =  2(1-r)A(p)\int_{-\infty}^{0^{-}} dx_1\; \chi_p(x_1-x_0)  + r  e^{-p g(x_r-x_0)} \widetilde{Q}_r(x_r,p) \;. \label{LT_FPE_norm}
\end{multline}
Note that the limits of the integral term  on the  l.h.s of \eqref{LT_FPE_norm} are from $0$ to $\infty$, which is different from the limits for a convolution which would be from $0$ to $z$. This prevents us from directly using the convolution theorem for Laplace transforms for solving this integral equation. Equation \eqref{LT_FPE_norm}, is now of the form of a Wiener-Hopf integral equation that, in principle, can be solved for $\widetilde{Q}_r(x_0,p)$ using the recipe provided in \ref{appendix_WH}. However, the  general solution of \eqref{LT_FPE_norm}, for general  resetting cost $g(x_r-x_0)$, is not in a tractable closed form. Nevertheless, we  obtain
tractable forms for  our chosen resetting costs:  a constant cost or linear cost per reset back to the origin. We can still let the jump distribution $f(\eta)$ and jump cost $h(\eta)$ be arbitrary.

\subsection{Simple solution cases of inhomogeneous Wiener-Hopf equation}
Let us summarise here some simple cases of the inhomogeneous Wiener-Hopf equation.
We consider the Wiener-Hopf  equation  in the form
\begin{align}
    \psi(z,p) -B(p)\int_0^\infty dz^\prime \psi(z^\prime,p) \chi(z-z^\prime) = J(z,p)\;, \label{WH1}   
\end{align}
where $\psi(z,p)$ is the function we seek a solution for, $\chi(\eta)$ is symmetric and normalised, and $J(z,p)$ is the inhomogeneous term.

The solution  of \eqref{WH1}, detailed in \ref{appendix_WH}, is given in terms of the Laplace transform 
\begin{equation}
    \widetilde \psi(\lambda,p) = \int_0^\infty dz\, e^{-\lambda z} \psi(z,p)\;. \label{psilambdadef}
\end{equation}
 In principle, the Laplace transform \eqref{psilambdadef} can be inverted to obtain $\psi(z,p)$ however it is simplest
to obtain an expression for $\psi(0,p)$  by taking $\lambda \to \infty$
and using Watson's lemma
\begin{equation}
    \lim_{\lambda \to \infty} \lambda \widetilde \psi(\lambda,p) = \psi(0,p)\;. \label{Wl}
\end{equation}

We also define the function
\begin{align}
    \phi(\lambda,p) = \exp\left[-\frac{\lambda}{\pi} \int_0^\infty dk\; \frac{\ln(1-B(p) \widehat{\chi}(k))}{\lambda^2 + k^2} \right]\;. \label{phi_expression}
\end{align}
where $B(p)$ is the function appearing in the l.h.s of \eqref{WH1} and
 \begin{align}
 \widehat{\chi}(k) = \int_{-\infty}^{\infty}d\eta\;e^{ik\eta}\chi(\eta)\;.  
\end{align}

We now enumerate the relevant known solution cases that we  will require.
\begin{enumerate}
    \item  In this case the inhomogeneous term is constant with respect to $z$:
    \begin{equation}
        J(z,p)=j(p)\;.\label{Jcon}
    \end{equation}
    The Laplace transform \eqref{psilambdadef} is given by
\begin{equation}
 \widetilde \psi(\lambda,p) =  \frac{j(p)}{\lambda} \frac{\phi(\lambda,p)}{\sqrt{1-B(p)}} \label{psi1}
\end{equation}
and we obtain, using  \eqref{Wl} and $\phi(\infty,p) =1$,
\begin{equation}
 \psi(0,p) =  \frac{j(p)}{\sqrt{1-B(p)}}\;.
 \end{equation}
\item In this case the inhomogeneous term decays exponentially with $z$
\begin{equation}
J(z,p)=j(p)e^{-pz}\;.\label{Jexp}
\end{equation}
The Laplace transform \eqref{psilambdadef} is given by
\begin{equation}
 \widetilde \psi(\lambda,p) =  \frac{j(p)}{\lambda+p} \phi(\lambda,p) \phi(p,p) \label{psi_exp_sol}
\end{equation}
and we obtain
\begin{equation}
 \psi(0,p) =  j(p)\phi(p,p)\;.
 \end{equation}
\item In this case the inhomogeneous term is of the following integral form: 
\begin{equation}
  J(z,p) = b(p) \int_{-\infty}^{-z} d\eta\,f(\eta)\;.  \label{Jint}
\end{equation} The Laplace transform \eqref{psilambdadef} is given by
 \begin{equation}
 \widetilde \psi(\lambda,p) =  \frac{b(p)}{B(p)} \frac{1}{\lambda} 
 \left[ 1-  \sqrt{1-B(p)} \phi(\lambda,p) \right] \label{psi3}
\end{equation}
and we obtain
\begin{equation}
 \psi(0,p) = \frac{b(p)}{B(p)} 
 \left[ 1- \sqrt{1-B(p)}  \right]\;.
 \end{equation} 
\end{enumerate}
\vspace*{1em}

 In order to  use the above results to construct the solution to \eqref{LT_FPE_norm}, we first note that in \eqref{LT_FPE_norm} the first inhomogeneous term on the right hand side may be written as
\begin{equation}
2(1-r)A(p)\int_{-\infty}^{0^-} dx_1\; \chi_p(x_1-x_0)
= 2(1-r)A(p)\int_{-\infty}^{-x_0^-} d\eta\; \chi_p(\eta)\;, \label{chi_property}
\end{equation}
which is a term of the form (iii).
The second inhomogeneous term in \eqref{LT_FPE_norm}  depends on the choice of
 $g(x_r-x_0)$, the cost of  a reset:  for $g(x_r-x_0)$  a constant, it is of the form (i) and for $g(x_r-x_0)$   linear
 it is of the form (ii).
  In the following two sections \ref{sec:const_cost} and \ref{sec:lin_cost}  we use the above results to construct the solution to \eqref{LT_FPE_norm} for  these choices of the resetting cost.

\section{Constant Cost per Reset} \label{sec:const_cost}

If we assume a constant cost per reset of the form
\begin{align}
    g(x_r-x_0) = c_r\;,
\end{align}
  making use of \eqref{chi_property}, we obtain from \eqref{LT_FPE_norm},
\begin{multline}
     \widetilde{Q}_r(x_0,p) -2(1-r)A(p)\int_0^\infty dx_1\; \widetilde{Q}_r(x_1,p)\chi_p(x_1-x_0)\\
     = 2(1-r)A(p)\int_{-\infty}^{-x_0^-} d\eta\; \chi_p(\eta) + r e^{-p c_r} \widetilde{Q}_r(x_r,p) \;, \label{LT_FPE_const}
\end{multline}
On the r.h.s. of \eqref{LT_FPE_const} we have two inhomogeneous terms. The first is of the form (iii) and the second is a constant term of the form (i). Since \eqref{LT_FPE_const} is a linear equation for $\widetilde{Q}_r(x_0,p)$ the solution is a superposition of the solutions for each of these inhomogeneous terms. Identifying $B(p) \to 2(1-r)A(p)$ in \eqref{WH1},   $ j(p) \to re^{-p c_r} \widetilde{Q}_r(x_r,p)$ in \eqref{psi1} and
$b(p) \to 2(1-r)A(p)$ in \eqref{psi3},  we can immediately write down the final solution for the Laplace transform of \eqref{LT_FPE_const} over $x_0$,  which we denote
\begin{equation}
   \widehat{Q}_r(\lambda,p) = \int_0^\infty dx_0\,e^{-\lambda x_0} \widetilde{Q}_r(x_0,p) 
\end{equation}
as
\begin{align}
    \widehat{Q}_r(\lambda,p) = \frac{1}{\lambda} \frac{r e^{-p c_r} \widetilde{Q}_r(x_r,p)}{\sqrt{1-2(1-r)A(p)}}\phi(\lambda,p) +  \frac{1}{\lambda} \left[ 1 - \sqrt{1-2(1-r)A(p)} \phi(\lambda,p) \right]\;. \label{gen_res_1}
\end{align}
where $A(p)$ is given in \eqref{A_def} and \eqref{phi_expression} becomes
\begin{align}
    \phi(\lambda,p) = \exp\left[-\frac{\lambda}{\pi} \int_0^\infty dk\; \frac{\ln(1-2(1-r)A(p) \widehat{\chi}_p(k))}{\lambda^2 + k^2} \right]\;. \label{phi_chi}
\end{align}

The expression \eqref{gen_res_1} simplifies significantly if we look at the solution for $x_0 = x_r = 0$. To obtain this particular case of the solution, we set $\lambda \to \infty$. Making use of Watson's lemma \eqref{Wl}, we finally obtain
\begin{align}
    \widetilde{Q}_r(0,p) = \frac{\sqrt{1-2A(p)(1-r)}-1+2A(p)(1-r)}{\sqrt{1-2A(p)(1-r)}-re^{-pc_r}}\quad \;. \label{no_reset_cost_sol}
\end{align}
Equation \eqref{no_reset_cost_sol} is the main result of this section; it provides the solution for the Laplace transform of the total cost distribution for any jump length distribution and any jump cost in the case of a constant resetting cost. Note that setting $r = 0$ reduces \eqref{no_reset_cost_sol} to the previously obtained result for the system without resetting \cite{ACEDK14,BCP22,pozzoli23,MMV24},
\begin{align}
    \widetilde{Q}_0(0,p) = 1-\sqrt{1-2A(p)}\;.
\end{align}
We will study in more detail the total cost distribution obtained in \eqref{no_reset_cost_sol} using three typical examples for jump distributions and jump cost functions in Sections \ref{sec:lapalce}-\ref{sec:levy_flight}. But first we will obtain the general asymptotic behaviours expressed by the total cost distribution \eqref{no_reset_cost_sol} in Section~\ref{sec:asymptotics}.
\subsection{Asymptotic behaviour} \label{sec:asymptotics}
From \eqref{no_reset_cost_sol} we can deduce the asymptotic, large $C$ behaviour of  the distribution $Q(0,C)$.
The inverse Laplace transform is given by the Bromwich integral and is dominated from a contribution from the singularity furthest to the right (the singularity with the largest real part).

Inspecting \eqref{no_reset_cost_sol} we see that there is a pole at a real value $p^*$ given by 
\begin{equation}
2(1-r)A(p^*) = 1- r^2 \e^{-p^*c_r}\label{pole}
\end{equation}
There is also a branch point  at a real value $p_2$, given by 
\begin{equation}
2(1-r)A(p_2) = 1
\end{equation}
Now since $A(p)$ given by \eqref{A_def} is a decreasing function of $p$, we deduce that $p^*>p_2$ and the pole is the dominant contribution.
Moreover,  as  $A(0) = 1/2$,  we deduce $p^*<0$.

The other possible source of  a singularity is a singularity of $A(p)$  at $p=0$. This is manifested by nonanalytic terms in the expansion of $A(p)$ about $p=0$:
\begin{equation}
    A(p) = \frac{1}{2} +a(p)  -b_\mu p^\mu
    \label{Apexp}
\end{equation}
where 
\begin{equation}
    a(p) = -pa_1+\frac{p^2}{2} a_2 \ldots
\end{equation}
and $a_m$ is the $m^{\rm th}$ moment of the jump cost $h(\eta)$
\begin{equation}
 a_m =  \int_{0}^{\infty} d\eta\; h^m(\eta)f(\eta)\;. \label{a_def}
\end{equation}
Here, the  first non-analytic term in the expansion is $-b_\mu p^\mu$ where $\mu$ is non-integer and $a(p)$ contains the analytic terms $(-1)^m a_m$ where $1\leq m \leq  \lfloor \mu \rfloor$. For example
if $f(\eta) \sim |\eta|^{-(1+\mu)}$, then if $0<\mu<1$ then $a(p)=0$,
if  $1<\mu<2$ then $a(p)=-a_1p$ {\it etc}.
If  $A(p)$  is non-analytic at $p=0$ this singularity will dominate the inversion of \eqref{no_reset_cost_sol}, since $p^*<0$. If $A(p)$ is analytic at $p=0$, the pole at $p=p^*$ will give the dominant contribution.

Thus, if $A(p)$ is analytic around $p=0$ the asymptotic behaviour will be given by 
\begin{equation}
     Q_r(0,C) \sim \e^{p^*C} \lim_{p\to p^*}\left[
(p-p^*)\widetilde{Q}_r(0,p)\right]\label{nc_reset_asymptotic}
\end{equation}
{\it i.e.} an exponentially decaying distribution. 
Whereas, if $A(p)$ is non-analytic around $p=0$,
the asymptotic behaviour can be obtained from \eqref{no_reset_cost_sol}, by expanding for small values of $p$ using \eqref{Apexp}.
Plugging \eqref{Apexp} into \eqref{no_reset_cost_sol}, and expanding for small values of $p$, we obtain the leading non-analytic behaviour
\begin{align}
     \widetilde{Q}_r(0,p) = 1+ c(p) - \frac{2b_{\mu}(1+\sqrt{r})}{\sqrt{r}}p^{\mu} + \mathcal{O}\left( p^{\mu+1} \right)\;,
\end{align}
where $c(p$) indicates analytic terms i.e, $1+c(p)$ is a polynomial of order 
$\lfloor \mu \rfloor$.
Upon performing the inverse Laplace transform, the leading contribution comes from the term containing $p^\mu$ and  we obtain the leading order behaviour
\begin{align}
     Q_r(0,C) \sim -\frac{2b_{\mu}(1+\sqrt{r})}{\Gamma(-\mu)\sqrt{r}} C^{-(1+\mu)}\;. \label{Q_power_law}
\end{align}
\subsection{Laplace Jump Distribution with Linear Jump Cost} \label{sec:lapalce}
\begin{figure}
    \centering
    \includegraphics[width = 0.6\textwidth]{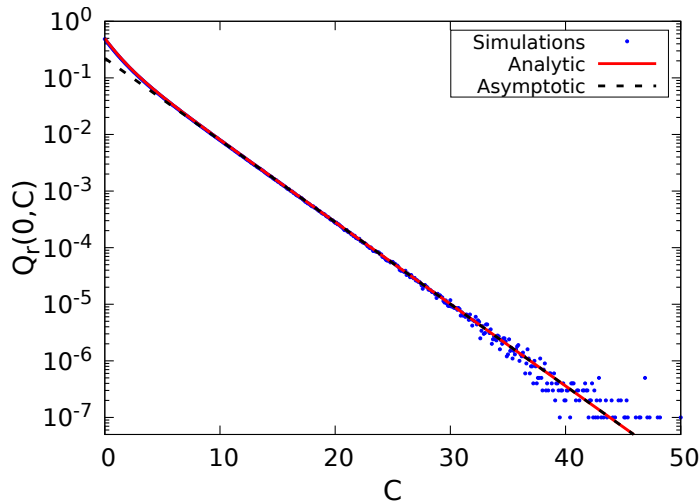}
    \caption{Distribution of cost obtained for Laplace jump distribution with linear jump cost  and cost-free resetting for $r = 0.5$. Note the logarithmic y-axis, indicating an exponential asymptotic behaviour.}
    \label{fig:Laplace_jump}
\end{figure}

First we consider the case which was exactly solved using a different method in Section~\ref{section_exact_solution},
\begin{align}
    f(\eta) &= \frac{1}{2}e^{-\abs{\eta}}\;,\\
    h(\eta) &= \abs{\eta}\;,\\
    g(\eta) &= 0\;,\\
    A(p) &= \int_0^{\infty}d\eta\;e^{-ph(\eta)}f(\eta) = \frac{1}{2(p+1)}\;. \label{Ap_laplace}
\end{align}
which upon plugging in \eqref{no_reset_cost_sol} gives,
\begin{align}
    \widetilde{Q}_r(0,p) = \frac{(p+1)(p+r)-p\sqrt{(p+1)(p+r)}}{(p+1)(p+r+pr)}\;,
\end{align}
which recovers \eqref{Laplace_cost_dist10}.
Numerically inverting the Laplace transform in \eqref{no_reset_cost_sol}, we obtain Fig.~\ref{fig:Laplace_jump}. The largest singularity of the above expression is given by the pole at $p = \frac{-r}{1+r}$, and we can compute the coefficient of the leading order term by computing the residue which gives,
\begin{align}
    Q_r(0,C) \sim \frac{2r^2}{(1+r)^2}e^{-\frac{r}{1+r}C}\;. \label{laplace_tail}
\end{align}

\subsection{Gaussian Jump Distribution with Quadratic Jump Cost} \label{sec:gaussian}

\begin{figure}
    \centering
    \includegraphics[width = 0.6\textwidth]{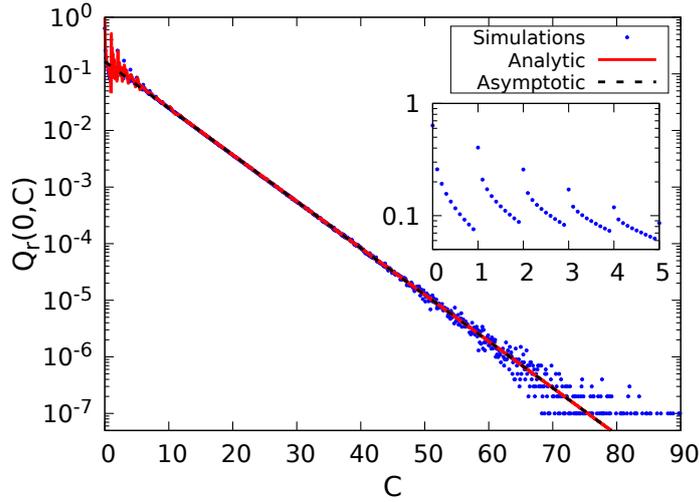}
    \caption{Distribution of cost obtained for Gaussian jump distribution with quadratic jump cost and unit reset cost for $r = 0.5$. Note the logarithmic y-axis, indicating an exponential asymptotic behaviour. The inset shows the spikes at integral cost values for the distribution, resulting from unit cost incurred at each reset.}
    \label{fig:Gaussian_jump}
\end{figure}

Now we consider the jump step sizes to be Gaussian distributed with quadratic jump cost,
\begin{align}
    f(\eta) &= \frac{1}{\sqrt{2 \pi}} e^{-\frac{\eta^2}{2}}\;,\\
    h(\eta) &= \eta^2\;,\\
    g(\eta) &= 1\;,\\
    A(p) &= \int_0^{\infty}d\eta\;e^{-ph(\eta)}f(\eta) = \frac{1}{2\sqrt{1+2p}}\;,
\end{align}
which upon using \eqref{no_reset_cost_sol} gives,
\begin{align}
    \widetilde{Q}_r(0,p) = \frac{-1+\sqrt{1+\omega}-\omega}{\sqrt{1+\omega}-re^{-p}}\;,
\end{align}
where $\omega = \frac{-1+r}{\sqrt{1+2p}}$. Inverting the above equation numerically, we obtain Fig.~\ref{fig:Gaussian_jump}. The largest pole is obtained numerically for $r=0.5$ by making use of \eqref{pole}, which gives $p^* = -0.1896\dots$. Plugging it in \eqref{nc_reset_asymptotic} we obtain the asymptotic result for large values of $C$ as $ Q_r(0,C) \sim 0.1649\,e^{-0.1897\,C}$. Another interesting feature is the presence of spikes as observed in the inset of Fig.~\ref{fig:Gaussian_jump} at integral values of the cost, which is due to the unit cost that each reset incurs. In this case, the spikes occur at integral values due to the choice of unit cost per reset. In general, spikes appear at integral multiples of the cost per reset $g(\eta)$. This can be clearly seen, for the case of no jump costs, by setting $h(\eta) = 0$, which corresponds to $A(p)=1/2$, in \eqref{no_reset_cost_sol}
\begin{equation}
\widetilde{Q}_r(0,p)=\frac{1-r^{1/2}}{1-r^{1/2} e^{-pc_r}}\;.
\end{equation}
Upon inverting one obtains
\begin{align}
    Q_r(0,C) = (1-\sqrt{r})\sum_{n=0}^{\infty} r^{n/2} \delta(C-n c_r)\;,
\end{align}
which is a superposition of delta-function spikes at integer multiples of the resetting cost $c_r$. 
The height of the spikes decrease with increasing values of $C$ as the probability of multiple resets before the first passage to the origin is rarer.
In Fig.~\ref{fig:Gaussian_jump} the spikes have more structure due to the presence of a jump cost.

\subsection{L\'evy Flight with Linear Jump Cost} \label{sec:levy_flight}
\begin{figure}
    \centering
    \includegraphics[width = 0.6\textwidth]{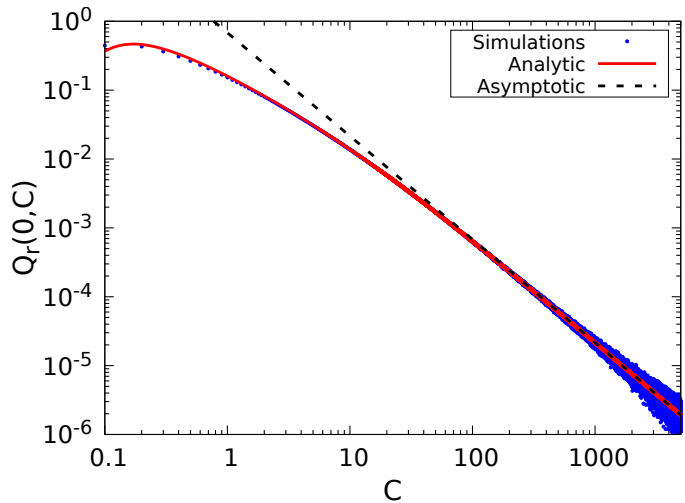}
    \caption{Distribution of cost obtained for L\'evy flights with linear jump cost  and cost-free resetting for $r = 0.5$. Note the logarithmic x and y axes, indicating a power-law asymptotic behaviour.}
    \label{fig:Levy_flight}
\end{figure}

The final  case we deal with is that of a L\'evy flight with L\'evy index $\mu = \frac{1}{2}$. We have the following jump distribution and costs,
\begin{align}
    f(\eta) &= \frac{1}{2\sqrt{4 \pi \abs{\eta}^3}}e^{-1/(4\abs{\eta})}\;,\\
    h(\eta) &= \abs{\eta}\;,\\
    g(\eta) &= 0\;,\\
    A(p) &= \int_0^{\infty}d\eta\;e^{-ph(\eta)}f(\eta) = \frac{1}{2}e^{-\sqrt{p}}\;. \label{Ap_levy}
\end{align}
 Plugging the above into \eqref{no_reset_cost_sol}  yields
\begin{align}
    \widetilde{Q}_r(0,p) = \frac{\sqrt{1-e^{-\sqrt p}(1-r)}-1+e^{-\sqrt p}(1-r)}{\sqrt{1-e^{-\sqrt p}(1-r)}-r}\quad \;. 
\end{align}
Numerical inversion of this expression along with simulations results are presented in Fig.~\ref{fig:Levy_flight}. We observe
a power-law decay of the cost distribution.

To understand the power-law decay we note that for L\'evy flights with $\mu =\frac{1}{2}$ with linear jump cost, the mean cost per jump defined by the integral $a_1$ \eqref{a_def} is divergent. So making use of \eqref{Q_power_law} and identifying $b_\mu = \frac{1}{2}$, we have the asymptotic behaviour for large $C$ as
\begin{align}
    Q_r(0,C) \sim \frac{1+\sqrt{r}}{2\sqrt{\pi r}}C^{-\frac{3}{2}}\;. \label{levy_tail}
\end{align}
This asymptotic expression is in good agreement with numerical results as illustrated in Fig.~\ref{fig:Levy_flight}.

We conclude that, unlike the previous cases where resetting changes the tail behaviour for large $C$ to an exponential decay, for power-law jump distributions with divergent mean cost per reset the power-law jumps still dominate over the resets and the large cost behaviour remains as a power-law decay.
\section{Linear Cost per Reset} \label{sec:lin_cost}
For the case of a linear cost per resetting where $x_r = 0$,
\begin{align}
    g(x_r - x_0) = x_0\;,
\end{align}
equation \eqref{LT_FPE} becomes,
\begin{align}
     \widetilde{Q}_r(x_0,p) - 2(1-r)A(p)\int_0^\infty dx_1\; \widetilde{Q}_r(x_1,p)&\chi_p(x_1-x_0)\nonumber \\&= \int_{-\infty}^{-x_0^-} d\eta \; \chi_p(\eta) + r e^{-p x_0} \widetilde{Q}_r(0,p) \;, \label{LT_FPE_norm_lin}
\end{align}
where again,
\begin{align}
    \chi_p(\eta) = \frac{e^{-ph(\eta)} f(\eta)}{2 A(p)} \quad \text{with} \quad A(p) = \int_0^{\infty}d\eta\;e^{-ph(\eta)}f(\eta)\;.
\end{align}
We now have an exponential inhomogeneous term of the form (ii) \eqref{Jexp} and another term of the form (iii) \eqref{Jint}. Making the change $B(p) \to 2(1-r)A(p)$ in \eqref{psi_exp_sol} and \eqref{psi3}, $ j(p) \to r\widetilde{Q}_r(0,p)$ in \eqref{psi_exp_sol} and $b(p) \to 2(1-r)A(p)$ in \eqref{psi3}, we can write down the solution to \eqref{LT_FPE_norm_lin} as
\begin{align}
    \widehat{Q}_r(\lambda,p) =  r\widetilde{Q}_r(0,p)\frac{\phi(\lambda,p)\phi(p,p)}{\lambda + p} + \frac{1}{\lambda}\left[ 1 - \sqrt{1-2(1-r)A(p)} \phi(\lambda,p) \right]\;, \label{gen_res_2}
\end{align}

Again making use of Watson's lemma \eqref{Wl}, we obtain the $x_0 = 0$ behaviour 
\begin{align}
    \widetilde{Q}_r(0,p) =  \lim_{\lambda \to \infty} \lambda \widehat{Q}_r(\lambda,p) = r\widetilde{Q}_r(0,p)\phi(p,p) + 1 -\sqrt{1-2(1-r)A(p)}\;,
\end{align}
which can be rearranged to obtain
\begin{align}
    \widetilde{Q}_r(0,p) = \frac{1 -\sqrt{1-2(1-r)A(p)}}{1-r\phi(p,p)}\;. \label{Q_linear}
\end{align}
where the expression \eqref{phi_expression} is explicitly
\begin{align}
    \phi(p,p) = \exp\left[-\frac{p}{\pi} \int_0^\infty dk\; \frac{\ln\left(1-(1-r) \int_{-\infty}^{\infty}d\eta\;e^{ik\eta}e^{-ph(\eta)}f(\eta)\right)}{p^2 + k^2} \right]\;. \label{phi_Ap}
\end{align}
The difficulty in making use of \eqref{Q_linear} lies in being able to evaluate the function $\phi(p,p)$ given by $\eqref{phi_Ap}$. However, results can be readily obtained by numerically computing the integral.

\subsection{Laplace Jump Distribution with Linear Jump Cost} \label{sec_gen_laplace_lin}

One of the cases which can be obtained in closed form is the Laplace jump distribution with linear jump considered in Section~\ref{section_exact_solution}. This serves as a useful check of our results. We have
\begin{align}
    f(\eta) &= \frac{1}{2}e^{-\abs{\eta}}\;,\\
    h(\eta) &= \abs{\eta}\;,\\
    g(\eta) &= \abs{\eta}\;,\\
    A(p) &= \int_0^{\infty}d\eta\;e^{-ph(\eta)}f(\eta) = \frac{1}{2(p+1)}\;.
\end{align}
To make use of \eqref{phi_Ap}, we need the Fourier transform
\begin{align}
\int_{-\infty}^{\infty}d\eta\;e^{ik\eta}\frac{1}{2}e^{-(p+1)\abs{\eta}} = \frac{(p+1)}{(p+1)^2+k^2}\;.  
\end{align}
Making use of the following integral identity,
\begin{align}
   \int_{0}^{\infty}dx\; &\frac{\log{(b+x^2)}}{a+x^2}  = \frac{\pi}{\sqrt{a}}\log\left( \sqrt{a}+\sqrt{b}\right)\;,
\end{align}
we can finally calculate from \eqref{phi_Ap},
\begin{align}
    \phi(p,p) = \frac{1+2p}{p+\sqrt{(1+p)(p+r)}}\;.
\end{align}
Plugging the above expression back into \eqref{Q_linear}, we obtain

\begin{align}
    \widetilde{Q}_r(0,p)= \frac{\sqrt{1+p}(r+ (2r+1)p)-p \sqrt{p+r}}{\sqrt{1+p} (4rp^2+(4r+1)p+r)}\;,\label{lin_cost_final}
\end{align}
which is the same as that obtained using the method of Section~\ref{section_exact_solution}. 

\subsection{L\'evy Flight with Linear Jump Cost} \label{sec_gen_laplace_lin_reset}
Another case which is of interest is that of L\'evy flights with linear jump and reset costs. However, \eqref{phi_Ap} is cumbersome to work with, but we can readily integrate it numerically after performing some calculations which simplify the expression and make use of \eqref{Q_linear} to obtain the total cost distribution in Laplace space. For L\'evy flights with linear costs we have 
\begin{align}
    f(\eta) &= \frac{1}{2\sqrt{4 \pi \abs{\eta}^3}}e^{-1/(4\abs{\eta})}\;,\\
    h(\eta) &= \abs{\eta}\;,\\
    g(\eta) &= \abs{\eta}\;,\\
    A(p) &= \int_0^{\infty}d\eta\;e^{-ph(\eta)}f(\eta) = \frac{1}{2}e^{-\sqrt{p}}\;. \label{Ap_levy_lin}
\end{align}
We also have the Fourier transform
\begin{align}
\int_{-\infty}^{\infty}d\eta\;e^{ik\eta}\frac{1}{2\sqrt{4 \pi \abs{\eta}^3}}e^{-p\abs{\eta}-1/(4\abs{\eta})} = \frac{1}{2} \left( e^{-\sqrt{p-ik}} + e^{-\sqrt{p+ik}} \right)\;.  
\end{align}
Equation \eqref{phi_Ap} can now be integrated numerically over $k$ and then plugged in \eqref{Q_linear} to obtain the Laplace transform of the total cost distribution. Finally, upon performing the inverse Laplace transform numerically, we obtain Fig. \ref{fig:Levy_flight_linear}. As observed in the case of L\'evy flights with constant resetting cost, here too the jump distribution dominates over the resetting behaviour and causes the asymptotic total cost distribution to decay as a power-law. 
The exponent is consistent with the value of $3/2$ found for the case of L\'evy flights with constant resetting cost in \eqref{levy_tail}. 
This is again indicative of the large cost behaviour being dominated by the power-law jump distribution.
\begin{figure}
    \centering
    \includegraphics[width = 0.6\textwidth]{Levy_linear_costs.pdf}
    \caption{Distribution of cost obtained for L\'evy flights with linear jump and reset costs for $r = 0.5$. The analytic behaviour is obtained by performing a numerical integration of \eqref{phi_Ap} and then further inverting \eqref{Q_linear} numerically. Note the logarithmic x and y axes, indicating a power-law asymptotic behaviour. {\blue The dashed line corresponds to a line of slope $-3/2$,  which indicates that the jump distribution dominates the total cost} }
    
    
    \label{fig:Levy_flight_linear}
\end{figure}

{\blue
\subsection{L\'evy Flight with only Linear Reset Cost}
Another simple but interesting case is that of L\'evy flights with linear reset costs only and no jump costs. As with the previous case of L\'evy flights with linear reset costs and jump costs, we start  from \eqref{Q_linear} and \eqref{phi_Ap}, use
\begin{align}
    f(\eta) &= \frac{1}{2\sqrt{4 \pi \abs{\eta}^3}}e^{-1/(4\abs{\eta})}\;,\\
    h(\eta) &= 0\;,\\
    g(\eta) &= \abs{\eta}\;,\\
    A(p) &= \int_0^{\infty}d\eta\;f(\eta) = \frac{1}{2}\; \label{Ap_levy_lin_2},
\end{align}
then resort to numerical integration of \eqref{phi_Ap}. 
We can calculate the Fourier transform required in \eqref{phi_Ap} as
\begin{align}
\int_{-\infty}^{\infty}d\eta\;e^{ik\eta}\frac{1}{2\sqrt{4 \pi \abs{\eta}^3}}e^{-1/(4\abs{\eta})} = e^{-\sqrt{\frac{k}{2}}}\cos\left( \sqrt{\frac{k}{2}} \right)\;.  
\end{align}
Numerically integrating \eqref{phi_Ap} over $k$ and then using \eqref{Q_linear}, we obtain the Laplace transform of the total cost distribution. Finally, upon performing the inverse Laplace transform numerically, we obtain Fig.~\ref{fig:Levy_flight_linear_reset_cost}.
\begin{figure}
    \centering
    \includegraphics[width = 0.6\textwidth]{Levy_linear_reset_cost.pdf}
    \caption{{\blue Distribution of cost obtained for L\'evy flights with  linear reset cost and no jump cost for $r = 0.5$. The analytic behaviour is obtained by performing a numerical integration of \eqref{phi_Ap} and then further inverting \eqref{Q_linear} numerically. Note the logarithmic x and y axes, indicating a power-law asymptotic behaviour. The dashed line corresponds to a line of slope $-3/2$, which is the expected power-law decay behaviour for the total cost.}}
    \label{fig:Levy_flight_linear_reset_cost}
\end{figure}
We observe from Fig.~\ref{fig:Levy_flight_linear_reset_cost} that the heavy-tailed jump distribution dominates over the resetting behaviour.
Thus in this case, as well as the case of 
Section \ref{sec_gen_laplace_lin_reset}, the result is consistent with the exponent $3/2$.
}

\section{Conclusions and Outlook} \label{sec:conclusion}
In this work we have extended the general framework to calculate the total cost distribution for discrete-time random walks up to first passage to the target presented in \cite{MMV24} to the case of random walks with resetting. We have obtained a general expression for the total cost distribution for constant resetting cost \eqref{gen_res_1} and linear resetting cost \eqref{gen_res_2}, which is valid for arbitrary jump distributions and jump cost functions. 

Furthermore, we observed that the dominant large cost behaviour is determined by the jump distribution. We saw that for the case of constant resetting cost, the presence of a power law tail for the jump distribution results in the large cost behaviour obtaining a power-law decay, which is similar to that of the case without resetting. In the absence of such a power-law tail, resetting dominates and the total cost function obtains an exponential decay at large values of total cost. 

It would be of significance to explore whether other reset cost functions, beyond constant and linear, could yield an analytically tractable solution from the Wiener-Hopf integral equation. 
{\blue Also for systems where there is a trade-off between jump and reset costs, a function which encapsulates the competition could be constructed and optimised in a similar fashion to  \cite{SBEM23}.}
Other interesting directions to pursue would be to calculate the penalty associated with overshooting the target barrier and compare the effect of resetting on it. Computing the results for the system in higher dimensions and multiple time step dependent cost functions would also be interesting.

 \section*{Acknowledgements}
For the purpose of open access, the authors have applied a Creative Commons Attribution (CC BY) licence to any Author Accepted Manuscript version arising from this submission.  JCS thanks the University of Edinburgh for the award of an EDC Scholarship.

\appendix
 
\section{Solution of Differential Equation for Exactly Solvable Case} \label{appendix_sol_DE}
From Section~\ref{sec:exact_sol} we have the differential equation for the total cost distribution for the Laplace jump distribution and linear costs as,
 \begin{align}
    \frac{d^2}{dx_0^2} \widetilde{Q}_r(x_0,p) = &(p c_j +1) (p c_j  + r) \widetilde{Q}_r(x_0,p) + r\Big[ (p c_r)^2 e^{-p c_r \abs{x_r - x_0}} \nonumber \\ &-(p c_j + 1)^2e^{-p c_r \abs{x_r-x_0}} - 2p c_r \delta(x_r - x_0) \Big]\widetilde{Q}_r(x_r,p)\;. \label{Laplace_DE_appendix}
\end{align}
 Due to the presence of a delta function in the inhomogeneous term in \eqref{Laplace_DE_appendix}, the differential equation has to be solved in two parts for $x_0 > x_r$ and $x_0<x_r$ and then joined together using the delta function jump condition at $x = x_r$. Solving for $x_0 >x_r$,
\begin{align}
    \frac{d^2}{dx_0^2} \widetilde{Q}_r(x_0,p) = &(p c_j +1) (p c_j  + r) \widetilde{Q}_r(x_0,p) + r\Big[ (p c_r)^2 e^{p c_r (x_r - x_0)} \nonumber \\ &-(p c_j + 1)^2e^{p c_r (x_r-x_0)} \Big]\widetilde{Q}_r(x_r,p)\;.\label{Laplace_DE_r}
\end{align}
Define $\psi(x_0,p) = e^{p c_r x_0}\widetilde{Q}_r(x_0,p)$. We then have
\begin{align}
    \frac{d^2}{dx_0^2}\widetilde{Q}_r(x_0,p)=e^{-p c_r x_0}\frac{d^2}{dx_0^2}\psi(x_0,p) - 2p c_r e^{-p c_r x_0} \frac{d}{dx_0}\psi(x_0,p) + (c_r p)^2 e^{-p c_r x_0}\psi(x_0,p)\;,
\end{align}
using which we can rewrite \eqref{Laplace_DE_r} as
\begin{align}
    \frac{d^2}{dx_0^2}\psi(x_0,p) -2p c_r \frac{d}{dx_0}\psi(x_0,p) +\left[(p c_r)^2 - (p c_j+r)(p c_j + 1)\right] \psi(x_0,p) \nonumber \\= r \left[(p c_r)^2 - (p c_j + 1)^2\right] \psi(x_r,p)\;. \label{inhom_DE}
\end{align}
The right hand side of \eqref{inhom_DE} is a constant inhomogeneous term, and \eqref{inhom_DE} can be converted to a homogeneous equation by defining
\begin{align}
    \psi(x_0,p) &= w(x_0,p) + a \quad \text{where} \quad  a = \frac{r \left[ (p c_r)^2 - (p c_j + 1)^2 \right]}{(p c_r)^2 - (p c_j+r)(p c_j + 1)} \psi(x_r,p)\;, \label{def_w}
\end{align}
which finally leads to
\begin{align}
    \frac{d^2}{dx_0^2}w(x_0,p) - 2p c_r \frac{d}{dx_0}w(x_0,p) + \left[(p c_r)^2 - (p c_j+r)(p c_j + 1)\right] w(x_0,p) = 0\;.
\end{align}
Assuming a solution of the form $w(x_0,p) = Ae^{\lambda x_0}$, we obtain the possible values of $\lambda$ as
\begin{align}
    \lambda = p c_r \pm \sqrt{(p c_j + r) (p c_j + 1)}\;.
\end{align}
So, finally making use of the definitions of $w(x_0,p)$ and $\psi(x_0,p)$, we obtain the solution for $x_0 > x_r$ (called the right solution, denoted by the super-script $r$) as
\begin{align}
    \widetilde{Q}^{(r)}_r(x_0,p) = A e^{-\sqrt{(p c_j + r) (p c_j + 1)}x_0} + \frac{r \left[ (p c_r)^2 - (p c_j + 1)^2 \right]}{(p c_r)^2 - (p c_j+r)(p c_j + 1)} e^{p c_r (x_r - x_0)} \widetilde{Q}_r(x_r,p)\;. \label{Qrr}
\end{align}
Note that we have not included the $e^{\sqrt{(p c_j + r) (p c_j + 1)}x_0}$ term for the right solution as we expect the solution to remain bounded as $x_0 \to \infty$.

Following a similar procedure, we can obtain a left solution for $x_0 < x_r$, which is given by
\begin{align}
    \widetilde{Q}^{(l)}_r(x_0,p) = &B e^{-\sqrt{(p c_j + r) (p c_j + 1)}x_0} + C e^{\sqrt{(p c_j + r) (p c_j + 1)}x_0} \nonumber \\ &+ \frac{r \left[ (p c_r)^2 - (p c_j + 1)^2 \right]}{(p c_r)^2 - (p c_j+r)(p c_j + 1)} e^{-p c_r (x_r - x_0)} \widetilde{Q}_r(x_r,p)\;. \label{Qrl}
\end{align}
Now, the last task is to find the coefficients $A,B$ and $C$, which can be determined using
\begin{itemize}
    \item[(i)] Continuity of the left and right solutions at $x_0 = x_r$.
    \item[(ii)] Discontinuity of the derivative at $x_0 = x_r$.
    \item[(iii)] The solution satisfying the original integral equation \eqref{Laplace_FPE}.
\end{itemize}
Requiring continuity at $x_0 = x_r$ gives
\begin{align}
    A-B = C e^{2\sqrt{(p c_j + r) (p c_j + 1)}x_r}\;,
\end{align}
and the derivative discontinuity at $x_0 = x_r$ gives,
\begin{align}
    C = \frac{r p c_r}{\sqrt{(p c_j + r)(p c_j + 1)}} \left[ 1 - \frac{(p c_r)^2 - (p c_j + 1)^2}{(p c_r)^2 - (p c_j+r)(p c_j + 1)} \right] e^{-\sqrt{(p c_j + r) (p c_j + 1)}x_r}\widetilde{Q}_r(x_r,p)\;.
\end{align}
Finally, we can solve for $A$ by plugging the solution into \eqref{Laplace_FPE}. Note that if we assume $x_0 > x_r$, we have to split the integration range into three intervals, $(0,x_r),(x_r,x_0)$ and $(x_0,\infty)$, with $\widetilde{Q}^{(l)}_r(x_0,p)$ for the first interval and $\widetilde{Q}^{(r)}_r(x_0,p)$ for the other two. The coefficient $A$ can be then obtained as 
\begin{align}
    A &= \frac{A_{\text{num}}}{A_{\text{den}}} \;,\\
    A_{\text{num}} &= e^{-\left(\mu +p c_r\right) x_r} (-1+r) \bigg\lbrace e^{\mu  x_r} \mu  \left(1+p c_j\right) \left[\widetilde{Q}_r(x_r,p) r \left(1+p c_j\right)-e^{p c_r x_r} \left(r+p
   c_j\right)\right]\nonumber \\&+ p \widetilde{Q}_r(x_r,p) r \left(1+p c_j\right) \Big[e^{p c_r x_r} \left(-1+\mu -p c_j\right)+e^{2 \mu  x_r+p c_r x_r} \left(1+\mu +p
   c_j\right)\nonumber \\&-e^{\mu  x_r} \mu\Big] c_r +e^{\left(\mu +p c_r\right) x_r} p^2 \mu  c_r^2\bigg\rbrace\\
   A_{\text{den}} &= \mu  \left(p c_j+\mu +1\right) \left(p \left(c_j \left(p c_j+r+1\right)-p c_r^2\right)+r\right)\;,
\end{align}
where $\mu = \sqrt{(1+pc_j)(r+pc_j)}$. Now using the expression for $A$, we now solve the expression \eqref{Qrr} self-consistently at $x_0=x_r$ to obtain the expression for $\widetilde{Q}_r(x_r,p)$. Finally, combining together $A,B,C$ and $\widetilde{Q}_r(x_r,p)$, along with \eqref{Qrr} and \eqref{Qrl}, we obtain the complete solution for the differential equation as presented in the supplemental Mathematica notebook \cite{DataShare}.

\section{Wiener-Hopf Integral Equation} \label{appendix_WH}
An equation of the form
\begin{align}
    \psi(z,p) -B(p)\int_0^\infty dz^\prime \psi(z^\prime,p) \chi(z-z^\prime) = J(z,p)\;, \label{WH_eq}   
\end{align}
is called a Wiener-Hopf integral equation. This appendix gives a recipe to solve the Wiener-Hopf problem. For more details we refer the reader to \cite{MCZ06,Majumdar10,Ivanov94}.

If $\chi(\eta)$ is continuous, symmetric and normalized to unity, \eqref{WH_eq} has a solution given in terms of the Green's function,
\begin{align}
    \psi(z,p) = \int_0^\infty dz_1\; G(z,z_1,p)J(z_1,p)\;,
\end{align}
where there exists an expression for $G(z,z_1,p)$ in terms of the double Laplace transform,
\begin{align}
    \widetilde{G}(\lambda,\lambda_1,p) = \int_0^\infty dz\;e^{-\lambda z} \int_0^\infty dz_1\;e^{-\lambda_1 z_1} G(z,z_1,p)\;,
\end{align}
which is given by,
\begin{align}
    \widetilde{G}(\lambda,\lambda_1,p) = \frac{\phi(\lambda,p)\phi(\lambda_1,p)}{\lambda + \lambda_1}\;. \label{LT_G}
\end{align}
Finally, the expression for $\phi(\lambda,p)$ is given by,
\begin{align}
    \phi(\lambda,p) = \exp\left[-\frac{\lambda}{\pi} \int_0^\infty dk\; \frac{\ln(1-B(p) \widehat{\chi}(k))}{\lambda^2 + k^2} \right]\;, \label{GF_survival}
\end{align}
where $\widehat{\chi}(k) = \int_{-\infty}^{\infty} dz\;e^{ikz}\chi(z)$ is the Fourier transform of $\chi(z)$.

The above set of steps are quite difficult to perform in general and extract useful information out of the solution, but the solution can be obtained in a relatively tractable form for a few special cases of $J(z,p)$.

\subsection{Constant $J(z,p)$: $J(z,p) = j(p)$}
If the inhomogenous term in \eqref{WH_eq} is a constant, then the Laplace transform of the solution, $\widetilde{\psi}(\lambda,s)$ can be expressed as
\begin{align}
    \widetilde{\psi}(\lambda,p) = \int_0^\infty dz\;e^{-\lambda z} \int_0^\infty dz_1\; j(p)G(z,z_1,p) = j(p) \widetilde{G}(\lambda,0,p) \;.
\end{align}
From \eqref{LT_G}, we obtain,
\begin{align}
    \widetilde{\psi}(\lambda,p) = \frac{j(p)}{\lambda} \phi(\lambda,p)\phi(0,p)\;,
\end{align}
and $\phi(0,p)$ can be expressed as \cite{MCZ06}
\begin{align}
    \phi(0,p) = \frac{1}{\sqrt{1-B(p)}}\;,
\end{align}
which leads to the final solution
\begin{align}
    \widetilde{\psi}(\lambda,p) = \frac{j(p)}{\lambda \sqrt{1-B(p)}} \phi(\lambda,p)\;. \label{WH_const_sol}
\end{align}
\subsection{Exponential $J(z,p): J(z,p) = e^{-pz}$} \label{appendix_WH_exp}
If the inhomogeneous term in \eqref{WH_eq} has an exponential form, then we can identify the integral over the inhomogeneous term as a Laplace transform and write,
\begin{align}
    \widetilde{\psi}(\lambda,p) = \int_0^\infty dz\;e^{-\lambda z} \int_0^\infty dz_1\; e^{-pz_1}G(z,z_1,p) = \widetilde{G}(\lambda,p,p) \;.
\end{align}
From \eqref{LT_G}, we get
\begin{align}
    \widetilde{G}(\lambda,p,p) = \frac{\phi(\lambda,p)\phi(p,p)}{\lambda + p}\;,
\end{align}
with $\phi(\lambda,p)$ given by \eqref{GF_survival}.

So we obtain the final solution as
\begin{align}
    \widetilde{\psi}(\lambda,s) = \frac{\phi(\lambda,p)\phi(p,p)}{\lambda + p}\;.
\end{align}

\subsection{Integral of form $J(z,p) = b(p) \int_{-\infty}^{-z} d\eta\,\chi(\eta)$} \label{appendix_WH_int}

Consider a linear transformation of the form
\begin{align}
    \psi_0(z,p) = a(p) \psi_1(z,p) - b(p)\;, \label{lin_trans}
\end{align}
where $\psi_0(z,p)$ satisfies,
\begin{align}
    \psi_0(z,p) -B(p)\int_0^\infty dz^\prime \psi_0(z^\prime,p) \chi(z-z^\prime) = j(p)\;.  
\end{align}
Using \eqref{lin_trans}, we get
\begin{align}
    a(p) \psi_1(z,p) - b(p) - B(p)\int_0^\infty dz^\prime \left[ a(p) \psi_1(z,p) - b(p) \right] \chi(z-z^\prime) = j(p)\;,
\end{align}
which can be rearranged to obtain
\begin{align}
    \psi_1(z,p) -B(p)\int_0^\infty dz^\prime \psi_1(z^\prime,p) &\chi(z-z^\prime) \nonumber \\= &\frac{b(p)}{a(p)}+\frac{j(p)}{a(p)}-\frac{B(p)b(p)}{a(p)}+\frac{B(p)b(p)}{a(p)} \int_{-\infty}^{-z} d\eta\,\chi(\eta)\;, \label{rearrange}
\end{align}
 where to obtain \eqref{rearrange}, we have made use of the symmetry and normalization of $\chi(\eta)$, which gives
\begin{align}
    \int_0^{\infty}dz^{\prime} \chi(z-z^\prime) = 1-\int_{-\infty}^{-z} d\eta\,\chi(\eta)\;.
\end{align}

Now choose $a(p) = B(p)$ and $j(p) = -b(p)+B(p)b(p)$, which leads us to
\begin{align}
    \psi_1(z,p) - B(p)\int_0^\infty dz^\prime \psi_1(z^\prime,p) \chi(z-z^\prime) = b(p)\int_{-\infty}^{-z} d\eta\,\chi(\eta)\;.
\end{align}
Now taking the Laplace transform of \eqref{lin_trans} and rearranging, we obtain
\begin{align}
    \widetilde{\psi}_1(\lambda,p) = \frac{1}{B(p)}\left( \widetilde{\psi}_0(\lambda,p)+\frac{b(p)}{\lambda} \right).
\end{align}
Finally making use of the solution to the Wiener-Hopf equation with the a constant cost \eqref{WH_const_sol} with $j(p) = -b(p)+B(p)b(p)$, we obtain the solution to the integral form of inhomogeneous term as,
\begin{align}
    \widetilde{\psi}_1(\lambda,p) = \frac{b(p)}{B(p)} \frac{1}{\lambda} \left[1- \phi(\lambda,p)\sqrt{1-B(p)}\right]\;.
\end{align}

\section{Simulation Details} \label{appendix_simulation}
To perform the simulation we first need to generate random variables according to the given jump length distribution, $f(x)$. This is obtained using an inverse transform sampling algorithm. We generate a uniform random number, $u \in [0,1]$ and obtain the jump length $x$ with the required distribution $f(x)$ by solving,
\begin{align}
    \int_{-\infty}^{x} d\eta\;f(\eta) = u\;.
\end{align}
We then sum up the costs up to the first passage time using the given jump cost function, $h(x)$ and reset cost function $g(x)$ and bin the results obtained appropriately. We sample over $10^8$ samples for each of the plots obtained in the paper. The program used for generating the total cost distribution is provided in \cite{DataShare}.
 \section*{References}

\bibliography{main}

\end{document}